\documentstyle[12pt]{article}

\newcommand{\nc}{\newcommand}
\nc{\beq}{\begin{equation}}
\nc{\eeq}{\end{equation}}
\nc{\beqa}{\begin{eqnarray}}
\nc{\eeqa}{\end{eqnarray}}

\input epsf
\newwrite\ffile\global\newcount\figno \global\figno=1
\def\writedefs{\immediate\openout\lfile=labeldefs.tmp \def\writedef##1{%
\immediate\write\lfile{\string\def\string##1\rightbracket}}}
\def\writestoppt{}\def\writedef#1{}
\def\figin{\epsfcheck\figin}\def\figins{\epsfcheck\figins}
\def\epsfcheck{\ifx\epsfbox\UnDeFiNeD
\message{(NO epsf.tex, FIGURES WILL BE IGNORED)}
\gdef\figin##1{\vskip2in}\gdef\figins##1{\hskip.5in}
\else\message{(FIGURES WILL BE INCLUDED)}%
\gdef\figin##1{##1}\gdef\figins##1{##1}\fi}
\def\figinsert{}
\def\ifig#1#2#3{\xdef#1{fig.~\the\figno}
\writedef{#1\leftbracket fig.\noexpand~\the\figno}%
\figinsert\figin{\centerline{#3}}\medskip\centerline{\vbox{\baselineskip12pt
\advance\hsize by -1truein\center\footnotesize{  Fig.~\the\figno.} #2}}
\bigskip\endinsert\global\advance\figno by1}
\def\footnotefont{}\def\endinsert{}

\begin{document}

\input epsf
\newwrite\ffile\global\newcount\figno \global\figno=1
\def\writedefs{\immediate\openout\lfile=labeldefs.tmp \def\writedef##1{%
\immediate\write\lfile{\string\def\string##1\rightbracket}}}
\def\writestoppt{}\def\writedef#1{}

\def\figin{\epsfcheck\figin}\def\figins{\epsfcheck\figins}
\def\epsfcheck{\ifx\epsfbox\UnDeFiNeD
\message{(NO epsf.tex, FIGURES WILL BE IGNORED)}
\gdef\figin##1{\vskip2in}\gdef\figins##1{\hskip.5in}
\else\message{(FIGURES WILL BE INCLUDED)}%
\gdef\figin##1{##1}\gdef\figins##1{##1}\fi}

\def\figinsert{}
\def\ifig#1#2#3{\xdef#1{fig.~\the\figno}
\writedef{#1\leftbracket fig.\noexpand~\the\figno}%
\figinsert\figin{\centerline{#3}}\medskip\centerline{\vbox{\baselineskip12pt
\advance\hsize by -1truein\center\footnotesize{  Fig.~\the\figno.} #2}}
\bigskip\endinsert\global\advance\figno by1}
\def\footnotefont{}\def\endinsert{}

\title{\Large{\bf 
Quark Condensates in Non-supersymmetric MQCD}}

\author{  
Nick Evans\thanks{nevans@physics.bu.edu} 
\\ 
{\small Department of Physics, Boston University, Boston, MA 02215} \\ \\
}

\date{}

\maketitle

\begin{picture}(0,0)(0,0)
\put(350,185){BUHEP-98-03}
\end{picture}
\vspace{-24pt}

\begin{abstract}
A set of non-supersymmetric minimal area embeddings of an M-theory
5-brane are considered. The field theories on the surface of the 5-brane
have the field content of N=2 SQCD with fundamental representation
matter fields. By suitable choice of curve parameters the N=2 and N=1
superpartners may be decoupled leaving a semi-classical approximation to
QCD with massive quarks.
As supersymmetry
breaking is introduced a quark condensate grows breaking the low energy
$Z_{F}$ flavour symmetry. At $\theta =$ (odd) $\pi$ spontaneous CP violation
is observed consistent with that of the QCD chiral lagrangian.
\end{abstract}

\newpage
\section{Introduction}

The massless states living on the surfaces of D-branes (for a review see
\cite{1}) in type IIA string
theory correspond to fields in a D+1 dimensional gauge theory
on the branes surface \cite{2}. D-brane constructions can therefore be used to
engineer field theories. In the string theory perturbative techiques can
be used to identify the effective field theory, and gauge theories in 4
dimensions have been realized with N=2 and N=1 \cite{3}-\cite{9} and N=0
\cite{4}\cite{6}\cite{9}-\cite{11} supersymmetry.  
To understand the strong dynamics of these gauge
theories one must study the IR properties of the brane
constructions. Within the context of string theory this is a difficult
problem since it involves the understanding of the cores of branes where
the dilaton vev blows up. 
Perturbative type IIA string theory is though the zero radius limit
of 11 dimensional M-theory. 
By considering related configurations in
theories where the radius of the eleventh dimension is brought up from
zero a continuous understanding of the interplay between D4 and NS5
branes may be obtained since they correspond to the same M5 brane in
some places wrapped on $x^{10}$ \cite{5}. 
A semi-classical approximation may then
be obtained for the IR behaviour of the configurations by making a
minimal area embedding of the M5 brane. These configurations have field
theories on their surfaces with the particle content of the string
theory field theories but with in addition extra Kaluza Klein modes with
masses of order the scale at which strong coupling dynamics sets in
\cite{6}. One
hopes that by moving to strong coupling in M-theory the low energy
dynamics remains in the same universality class as the theory on the
type IIA branes. For N=2 and N=1 theories in 4 dimensions this is born
out by the recovery of the field theory solution from the brane
configurations. For the one non-supersymmetric M5 brane minimal area 
embedding studied to date \cite{6}\cite{9}-\cite{11}, 
which has a low energy field theory described
by softly broken N=2 SQCD \cite{10}, confinement and the 
existence of QCD string solutions
has been observed \cite{6}\cite{9}. 
The calculation of the string tension in the brane
configuration \cite{9} and the field theory \cite{10} 
do not agree but this is not
surprising since the loss of supersymmetry relaxes the symmetry
constraints on the semi-classical approximation. Nevertheless one may
hope to deduce qualitative behaviour from the approximation. A recent
analysis \cite{11} of these configurations has demonstrated the non-trivial
spectral flow in the theories with changing theta angle previously
observed in field theory \cite{15}  with perturbing soft supersymmetry
operators \cite{15}-\cite{14}. 
The novel property of the string theory construction is
though that there is no restriction to small supersymmetry breaking
operators and one may decouple all superpartners.

In this paper we extend these previous analyses by introducing a
non-supersymmetric minimal area embedding of an M5 brane that
corresponds to a IIA configuration with semi-infinite D4 branes
that contribute quark flavours in the fundamental representation of the
gauge group. As supersymmetry breaking is switched on it is consistent
to interpret the extra parameter with a quark condensate. In the limit of
decoupling all superpartners one observes that the $Z_{F}$ remnant of
the quarks axial symmetry is broken by the IR dynamics again
demonstrating the existence of a quark condensate. Following
\cite{11} one observes non-trivial spectral flow as the bare theta angle
is changed in the theory. At $\theta =$ (odd) $\pi$ there are phase
transitions as two of $F$ discrete vacua interchange. This is
precisely the behaviour of the QCD chiral lagrangian  \cite{16}.

\section{Non-supersymmetric Type IIA Configurations}

Let us begin from the standard type IIA brane construction that realizes
a 4 dimensional N=2 $SU(N) \times SU(F)$ gauge theory 
with matter fields in the $(N, \bar{F})$ representation. From left
to right in the $x^6$ direction the branes are 

\beq \label{N=2config}
\begin{tabular}{|c|c|c|c|c|c|c|c|c|}
\hline
 & $\#$ & $R^4$ & $x^4$ & $x^5$ &  $x^6$ & $x^7$ & $x^8$ & $x^9$ \\
\hline 
NS5 & 1 & $-$ & $-$ & $-$  &  $\bullet$ & $\bullet$ & $\bullet$ & $\bullet$ \\
\hline
D4 & $N$ & $-$  & $\bullet$ &  $\bullet$ &  $[-]$ & $\bullet$ & 
                              $\bullet$ & $\bullet$ \\
\hline
NS5' & 1 & $-$ & $-$ & $-$  &  $\bullet$ & $\bullet$ & $\bullet$ & $\bullet$ \\
\hline
D4' & $F$ & $-$  & $\bullet$ &  $\bullet$ &  $[-]$ & $\bullet$ & 
                              $\bullet$ & $\bullet$ \\
\hline
$NS5^{''}$ & 1 & $-$ & $-$ & $-$  &  $\bullet$ & $\bullet$ & $\bullet$ & $\bullet$ \\
\hline
\end{tabular} 
\eeq

$R^4$ is the space $x^0-x^3$. A dash $-$ represents a
direction along a brane's world wolume while a dot $\bullet$ is
transverse. For the special case of the D4-branes' $x^6$ direction,
where the world volume is a finite interval corresponding to their
suspension between two NS5 branes at different values of $x^6$, 
we use the symbol $[-]$. 
The field theory exists on scales much greater 
than the $L_6$ distance between the NS5 branes with  the fourth
space like direction of the D4-branes generating  the couplings of the
gauge groups in the  effective 4D theory.

The $U(1)_R$ and $SU(2)_R$ symmetries of the N=2 field theory are
manifest in the brane picture. They correspond to isometries of the
configuration; an SO(2) in the  $x^4,x^5$ directions and an SO(3) in the
$x^7, x^8, x^9$ directions.

The matter multiplets are provided by $4-4'$ strings. Alternative
realizations of the matter fields are possible. For example 
the NS5$^{''}$ brane may be replaced with a D6 brane \cite{3}. 
The matter fields
remain but the D4-D6 boundary conditions freeze the flavour gauge field
and by supersymmetry its superpartners (when the configuration is broken
to N=1 supersymmetry the adjoint matter field of the frozen N=2 gauge
multiplet begins to propagate and is the ``meson'' of dual SQCD). The
flavour gauge multiplet may also be frozen by taking the NS5$^{''}$ to
infinity leaving a semi-infinite D4$'$ brane \cite{1}. 

Supersymmetry may be broken in the configuration in a number of ways
(see \cite{4}\cite{6} \cite{9}-\cite{11} for previous discussions). 
The NS5
branes may be rotated
in the  $x^4,x^5,x^7, x^8, x^9$ space (rotations from the N=2
configuration into the $x^6$
direction cause the NS5 branes to cross changing the topology of the
configuration in such a way that it can no longer be easily identified
with a field theory). Only the rotations of two of the three NS5 branes
correspond to changing the parameters of the field theory since a
rotation of the third can be reproduced by rotations of the other two
and a rotation of the whole configuration. There are six parameters
describing the rotations of each of the two NS5 branes (say the end two)
(SO(5)/SO(2)/SO(3)). 

Supersymmetry may also be broken by forcing the D4 and D4$'$ branes to lie at
angles to each other. In the supersymmetric configuration for the D4$'$ to
end on the NS5$'$ the $x^6-x^9$ coordinates of these branes must be shared
at the point where they meet. There is a choice for the $x^4-x^5$
coordinates since the NS5$'$ lies in those directions. This choice
determines the minimal length of the $4-4'$ strings and hence the mass of
the matter multiplet (for the N=2 configurations a mass term and an
adjoint vev are indistinguishable in both the field theory and the brane
configuration corresponding to the D4s freedom to move in the $x^4-x^5$
directions, the precise identification with a mass term is therefore
valid for N=1 configurations). To preserve supersymmetry the $x^4-x^5,x^7-x^9$
coordinates of the point where the D4$'$ ends on the NS5$^{''}$ are then
completely determined by the choice of mass. If we are
willing to break supersymmetry this need
not be the case. In general we may make arbitrary choices for the four
point like coordinates of the NS5$^{''}$ so that the D4$'$ corresponds to
an arbitrary vector in that four dimensional space. Its position
relative to the D4 branes is then
described by three angles.

The positions of the branes in these configurations break supersymmetry
and hence we expect there to be supersymmetry breaking parameters introduced
in the low energy field theory lagrangian. These parameters must be the
supersymmetry breaking vevs of fields in the string theory since at tree
level there are no supersymmetry breaking parameters. The vevs occur as
parameters because the fluctuations of those fields are being neglected
in the field theory; such fields are spurions. They have a natural
interpretation in the brane configuration. The fields are those
describing the positions of the branes and their fluctuations are
neglected because the infinite branes are very massive. If though we
choose to include these fields in the field theory description they
occur subject to the stringent constraints of N=2
supersymmetry\cite{13}\cite{14}. 
The
spurions whose vevs correspond to the supersymmetry breaking parameters
must be the auxilliary fields of N=2 multiplets. This constraint is
sufficient to identify the spurions.

Field theoretically N=2 Yang Mills theory without matter is very
restrictive on where spurions may occur. 
The unique possibility for
lowest dimension operators is that
spurion fields occur as vector fields 
in the prepotential as ${\cal F} = (S_1 + i
S_2)A^2$ \cite{10}\cite{12}. 
The scalar spurion vevs generate the gauge coupling $\tau$.
These spurions are natural candidates to
correspond to the supersymmetry breaking induced by rotations of the NS5
branes. When we allow the auxilliary fields of the spurions to be
non-zero we obtain the tree level masses

\begin{eqnarray} \label{softUV}
&& \nonumber
-{N_c \over 8  \pi^2} Im \left( (F_1^*+iF_2^*) \psi_A^\alpha \psi_A^\alpha 
+ (F_1 + i F_2)
\lambda^\alpha \lambda^\alpha + i \sqrt{2}
(D_1 + i D_2) \psi_A^\alpha\lambda^\alpha \right) \\
&&
  - {N_c \over 4 \pi^2 Im (s_1+is_2)} \left((|F_1|^2 + D_1^2/2) Im(
a^\alpha)^2
+  (|F_2|^2 + D_2^2/2) Re
(a^\alpha)^2 \right. \\
&& \left. \right. \hspace{4cm} + \left. 
 (F_1 F_2^* + F_1^*F_2 + D_1D_2) Im(a^\alpha) Re(a^\alpha)
\right)\nonumber
\end{eqnarray}
A number of consistency checks support the identification \cite{10}. 
Switching on
any one of the six  independent real supersymmetry breakings in the field
theory leaves the same massless spectrum in the field theory as in the
brane picture when any one of the six independent rotations of the NS5
brane is performed. The field theory and brane configurations possess
the same sub-manifold of N=1 supersymmetric configurations.

With the introduction of matter fields in the field theory a single
extra spurion field is introduced associated with the quark mass. The
only possibility is to promote the mass to an N=2 vector multiplet
associated with $U(1)_B$ \cite{14}.  Switching on its
auxilliary field vevs induce the tree level supersymmetry breaking 
operators
\beq
2 Re( F_{M}  q \tilde{q}) + D_{M} \left( |q|^2 -
|\tilde{q}|^2 \right)
\eeq
These breakings are therefore natural candidates to play the role of the
breakings induced by the angles between the D4 and D4$'$ branes. Again a
number of consistency checks support this identification. There are
three independent real parameters in both the field theory and the brane
picture. The scalar masses in the field theory 
break $SU(2)_R$ but leave two  $U(1)_R$
symmetries of the supersymmetric theory intact. The scalar masses may 
always be
brought to diagonal form by an $SU(2)_R$ transformation that mixes $q$
and $\tilde{q}^*$. In the resulting basis there is an unbroken $U(1)$
subgroup of $SU(2)_R$.
In the brane picture the
D4$'$ branes lie at an angle in the $x^6-x^9$ directions breaking the
$SU(2)_R$ symmetry but leaving two $U(1)_R$ symmetries unbroken. 

The resulting field theory potential has an interesting dependence on
the quark mass. The potential is of the form
\beq
V \simeq D (|q|^2 - |\tilde{q}|^2)~ + ~m^2 (|q|^2 + |\tilde{q}|^2)~ + ~
(|q|^2 - |\tilde{q}|^2)^2
\eeq
For $m=0$ the theory has a moduli space where $(|q|^2 - |\tilde{q}|^2) =
-D/2$. For $m^2<D$ the theory has a unique vacuum with $\langle q \rangle
= 0$ and $\langle |\tilde{q}|^2 \rangle = (D-m^2)/2$. 
In both cases the colour gauge
group is broken to an $SU(N-F)$ subgroup. For $m^2>D$ there is a unique
vacuum at the origin of moduli space and the gauge groups are unbroken. 
This same behaviour can be seen in
the brane picture. Consider for example the N=1 configuration of Fig 1
in which we move the central NS5 brane in the $x^9$ direction to put the
D4 branes at an angle. There are two possible results after switching on
$D$. One is that the D4 branes remain connected as before. The other is
that F of the D4 branes disconnect from the central NS5 brane
breaking the surface field theory to an $SU(F) \times SU(N-F)$ gauge
group. Which of these is energetically prefered depends on the relative
lengths of the separation between the NS5s in the $v$ direction (the
field theory mass) and the separation of the NS5s in the $x^9$
direction (corresponding in the field theory to the size of $D$). 
From Fig 1 the higgs branch is prefered when 
\beq
m^2 < 2d^2 - 2LL' + 2 \sqrt{L^2 + d^2}~ \sqrt{L`^2 + d^2}
\eeq \vspace{0.5cm}

$\left. \right.$  \hspace{-1cm}\ifig\prtbdiag{}
{\epsfxsize16truecm\epsfbox{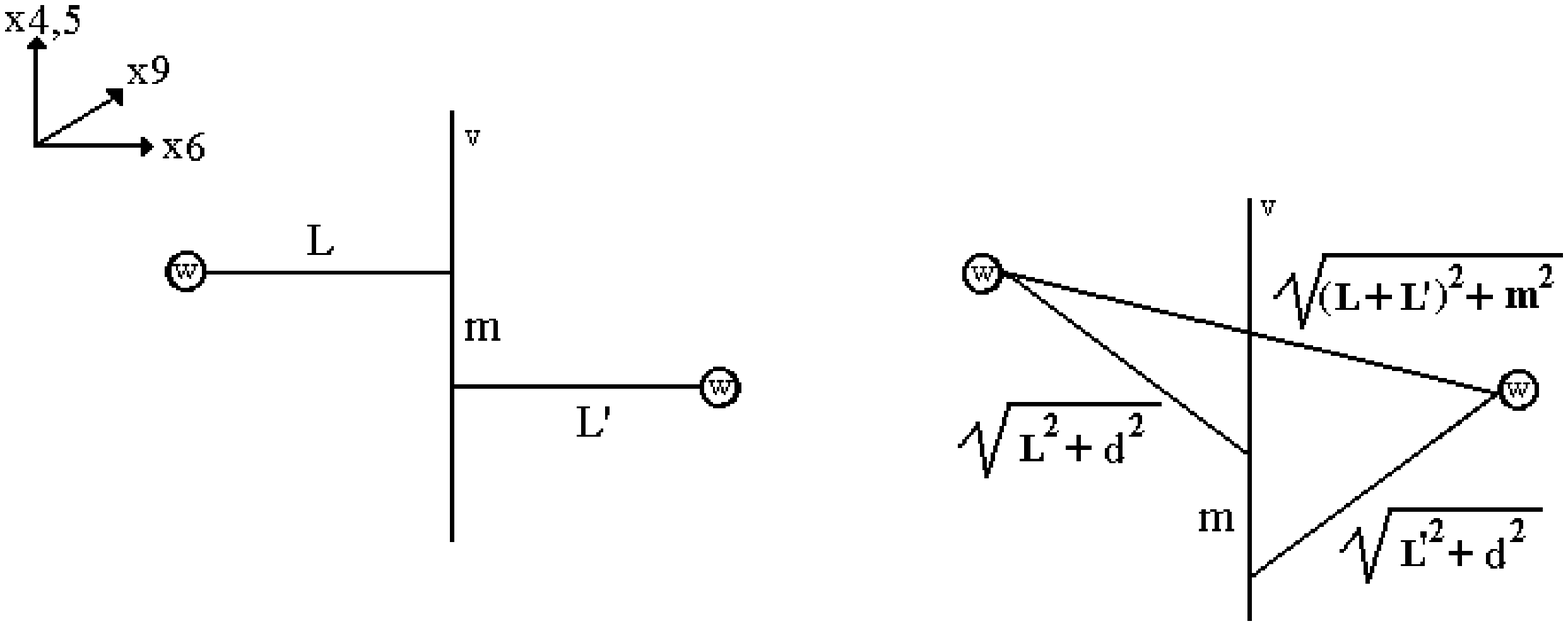}}  \vspace{-1.5cm}

\begin{center} Fig.1: An N=1 configuration with two NS5 branes in the
  $w$ direction and one in the $v$ direction. The same configuration after the
  central NS5 brane has been moved in the $x^9$ direction by
  displacement $d$, switching on 
  scalar mass soft breakings.\end{center}

Finally we note that the interpretation of the angles between D4 branes
as the expectation values of the auxilliary fields of the $U(1)_B$ gauge
multiplet is equivalent to the standard lore \cite{3} from N=1 theories that
motion of NS5 branes in the $x^9$ direction corresponds to switching on
a Fayet-Iliopoulos term for that field. 
Above we have simple considered the $U(1)_B$
field as a spurion rather than a propagating field (this is because, when
we move to M-theory configurations below, the classical $U(N)$ gauge
theory on the brane's surface is known to be broken to SU(N) with the
dynamics freezing the $U(1)$ field \cite{5}). 

The $SU(N) \times SU(F)$ brane configuration had six parameters
associated with the rotations of each of  
the end two NS5 branes which are
associated with the auxilliary field vevs of the two gauge coupling
spurions. The three parameters describing the angles between the D4 and
D4$'$ branes correspond to the auxilliary field vevs of the mass spurion. 
To study the dynamics of QCD with matter below we will take the
NS5$^{''}$ brane to infinity removing the six parameters associated with
its rotations and decoupling the flavour gauge multiplet. The resulting
theory has nine supersymmetry breaking parameters.

\section{IR Dynamics from M-theory}

To obtain the IR dynamics of the theory it is neccessary to study the
short distance behaviour of the brane configurations. In the string
theory the strongly coupled core of the NS5 brane prevents detailed
study of the NS5 D4 brane junction. Instead we move to M-theory
\cite{5}-\cite{7} by
increasing the $x^{10}$ dimension's radius, $R$, from zero. The NS5 and D4
branes are then aspects of the same M5 brane, wrapped in places around
the $x^{10}$ dimension. We may now perform a semi-classical minimal area
embedding of the surface to study the IR dynamics of the theory on the
brane's surface.

Following Witten \cite{6} we generalize the supersymmetric curves by the addition
of extra parameters. To find as many configurations as
possible we let the
configuration wallow in the full six dimensional space $\vec{X} 
= (x^4,..,x^9)$ and
$x^{10}$ which is picked out by its compact nature. The ansatz is

\begin{eqnarray}
\vec{X} & = & Re \left(\vec{p} z + \vec{q}/z + R\vec{r} \ln z +  
R \vec{s} \ln (z-m) \right)\nonumber \\
x^{10} & = & -N_c Im (\ln z) + N_f Im \ln (z-m)
\end{eqnarray}
$\vec{p}$ and $\vec{q}$ are complex 6 vectors, $\vec{r}$ and
$\vec{s}$
 real 6 vectors (this reality is required
to keep $\vec{X}$ single valued allowing a physical interpretation of the
configuration as a field theory \cite{11}). To enforce that the description is of
a minimal area embedding we require the vanishing of the two dimensional
energy-momentum tensor

\beq \label{cond}
T_{zz} = g_{ij} \partial_z X^i \partial_z X^j = 0
\eeq
where $g_{ij}$ is the background metric 

\beq
 ds^2 = \sum_{i,j=0}^9 \eta_{ij} dx^i dx^j + R^2 (dx^{10})^2
\eeq
Enforcing this condition leads to the constraints 

\beq
\begin{array}{c}
\vec{p}^2 = \vec{q}^2 = \vec{p}.\vec{r} = \vec{p}.
\vec{s} = \vec{q}.\vec{r} = \vec{q}.\vec{s} = 0
\\
\\
-2 \vec{p}.\vec{q} + R^2\vec{r}^2 = R^2N_c^2, \hspace{1cm}
\vec{r}.\vec{s} = -N_cN_f,\hspace{1cm}  \vec{
s}^2 = N_f^2  \end{array}
\eeq

There are 36 real parameters in the ansatz that are reduced by these
constraints and by using the background SO(6) isometries. We choose to
use 11 of the SO(6) degrees of freedom to fix parameters; this leaves an
$SO(2) \otimes SO(3)$ symmetry unfixed corresponding to leaving the
$U(1)_R \otimes SU(2)_R$ of the field theory unfixed. A further 2
parameters may be removed by scalings of $z$ (the parameter $m$ must
also be scaled). Finally the 7 complex 
constraints and 2 real constraints reduce the parameter count by a
further 16. We are left with 7 real parameters in $\vec{p},\vec{q},\vec{r}$ 
and $\vec{s}$ and 2 in $m$. 
This should be compared with
the field theory in which we argued that there were 2 associated with
the gauge coupling and theta angle, 2 with the mass parameter
and 9 with soft breakings. The mismatch
indicates that we do not have the most general set of field theories
(equivalently when we take the $ R \rightarrow 0$ limit in the M-theory
curves we do not recover the entire set of type IIA brane
configurations). This is not a surprise since the ansatz taken above is not
the most general solution to (\ref{cond}). 
The configurations above though are a subset of the non-susy field
theories with both gaugino and scalar masses and we proceed to analyse
those since they contain configurations with non-supersymmetric QCD with
matter in the IR.

A convenient parameterization of the curve is obtained as follows. We
use the constraint $\vec{p}^2=0$, the rescaling of $z$ ($m$ must also be
rescaled) and 8 of the SO(6)
isometry degrees of freedom (leaving 7 degrees of isometry freedom which
correctly corresponds to having fixed a single plane in 6-space leaving
an $SO(2) \times SO(4)$ isometry group) to set 

\beq \vec{p} = (1,-i,0,0,0,0). \eeq
The constraints $\vec{p}.\vec{s}=0$, $\vec{s}^2=N_f^2$ and 
3 of the remaining isometries may be used
to set

\beq \vec{s} = (0,0,N_f,0,0,0) \eeq
$\vec{p}.\vec{r}=0$, the $\vec{r}.\vec{s}$
constraints and a further 2 isometries may be used to set

\beq \vec{r} = (0,0,-N_c,0,0,a/R) \eeq

Note that we have used 2 of the 3 SO(3) or $SU(2)_R$ degrees of freedom
here. In the field theory we have therefore used $SU(2)_R$ to set two of
the  N=2 spurion components to zero. We are left with the isometries
corresponding to the two $U(1)_R$ symmetries of the N=1 configurations
unfixed.

The remaining constraints on $\vec{q}$ lead to a form 

\beq \vec{q} = ( \eta + \epsilon, -i \eta + i \epsilon, 0, \xi + \lambda, -i
\xi + i \lambda, 0) \eeq
and the constraints

\beq \begin{array}{c}
\eta \epsilon + \xi \lambda = 0 \\
\\
\epsilon = a^2/4 \end{array}
\eeq

The curve may then be written in terms of the more familiar variables
$v=(x^4+ix^5)$ and $w=(x^7+ix^8)$ as

\beq \begin{array}{c}
v = z + {\eta \over z} + {\bar{\epsilon} \over \bar{z}} \\
\\
w = {\xi \over z} + {\bar{\lambda} \over \bar{z}} 
\end{array} \eeq

The supersymmetric configurations with semi-infinite D4s can be
recovered by setting $a=0$. The constraints then set $\epsilon =0$ and
$\lambda = 0$. The curve is described in the
remaining space by $x^{9} = 0$ and setting $t=exp-(x^6/R+ix^{10})$ 

\beq 
t = \kappa z^{N}/(z-m)^{F}
\eeq

The constant $\kappa$ is undetermined by the minimal area embedding and
will be fixed to ensure the curve has the correct flow as quark masses
are taken to infinity and decoupled \cite{8}.
The curve has two $U(1)$ symmetries associated with rotation in the $v$
and $w$ planes which are broken by the parameters of the curve. The
symmetries may be restored by assigning the parameters spurious charges
\beq
\begin{array}{c|cccccccccc}
& v & w & t&  z& m& \eta & \epsilon & \xi & \lambda & \kappa\\
\hline
U(1)_v & 2 & 0 & 0 & 2 & 2 & 4 & 0 & 2 & 2 & 2(F-N)\\
U(1)_w & 0 & 2 & 0 & 0 & 0 & 0 & 0 & 2 & -2 & 0 \\
\end{array}
\eeq

\section{Supersymmetric Configurations}

We first review the supersymmetric theories desribed by the curve when
$a= \epsilon = \lambda =0$
\beq
v = z + {\eta \over z}, \hspace{1cm} w = {\xi \over z}, \hspace{1cm} t =
\kappa z^{N}/(z-m)^F
\eeq

From the string theory limit we know that the curve describes N=2 SQCD
broken to N=1 with an adjoint mass. If $\xi \rightarrow 0$ the curve
describes N=2 SQCD  at one point on moduli space and if $\eta \rightarrow
0$ the curve describes N=1 SQCD with the adjoint matter decoupled.
Furthermore the parameter $m$
plays the role of the field theory matter field mass term, $m_Q$. 
It is natural
therefore to associate the $U(1)_v$ and $U(1)_w$ symmetries
with the field theory $U(1)_R$ symmetries
\beq
\begin{array}{c|ccccccc}
& W & A& Q& \tilde{Q}& m_Q & m_A & \Lambda_{N=2} \\
\hline
U(1)_R & 1 & 2 & 0 & 0 & 2 & -2 & 2\\
U(1)_{R'} & 1 & 0 & 1 & 1 & 0 & 2 & 0\\
\end{array}
\eeq
where $m_Q$ is a common matter field mass, $m_A$ the adjoint field mass,
and $\Lambda_{N=2} \sim exp(2\pi i \tau/ ~(2N-F))$ the strong coupling
scale of the N=2 theory.
We can then make identifications between the field theory parameters and
brane configuration parameters. In particular we
may idenitfy $\xi$ with $m_A \Lambda_2^{b_0/N}m_Q^{F/N}$, $\eta$ with
$m_Q^{F/N} \Lambda_2^{b_0/N}$ and the
curve parameter $m$ with the field theory quark mass. 
 The constant $\kappa$ may now be
identified with $m_Q^{F-N}$ in order that as a single quark mass is
taken to infinity the curve correctly flows to the curve with $F
\rightarrow F-1$ \cite{8}.

While the curve's parameters in general break the two $U(1)$ symmetries,
asymptotically descrete subgroups are preserved. Asymptotically as
$\Lambda_{N=2} \rightarrow 0$ the
curve is given by
\begin{eqnarray}
z \rightarrow \infty & w=0 & t = v^{N-F} m_Q^{F-N} \nonumber\\
z \rightarrow 0 & v = 0 & t = \left( {m_A \over w}\right)^N 
\Lambda_{N=2}^{b_0} m_Q^{F-N}
\end{eqnarray}
The asymptotic curve leaves $U(1)_w$ unbroken (allowing $m_A$ to
transform spuriously) whilst only a $Z_{2N-F}$
subgroup of the $U(1)_v$ remains (allowing $m_A$ and $m_Q$ to
transform). This is precisely the effect of the
anomaly on the two $U(1)_R$ symmetries of the field theory.

\subsection{N=1 SQCD}

The N=1 curve with the adjoint matter field completely decoupled 
may be obtained by taking $\eta \rightarrow 0$ at fixed $\xi$
and is described by
\beq v = z, \hspace{1cm} w = {\xi \over z}, \hspace{1cm} t =
z^{N}m^{F-N}/(z-m)^{F} \eeq
The curve again has two $U(1)$ symmetries which correspond to
the field theory symmetries
\beq \label{v}
\begin{array}{c|ccccc}
& W & Q & \tilde{Q} & m_Q & \Lambda^{b_0}\\
\hline
U(1)_R & 1 & 0 & 0 & 2 & 2(N-F) \\
U(1)_{R'} & 1 & 1 & 1 & 0 & 2 N \\
\end{array}
\eeq
where $\Lambda = exp(2\pi i \tau/b_0)$ and $b_0 = 3N-F$. We may
make the identifications $\xi = \Lambda^{b_0/N} m_Q^{F/N}$ and $m=m_Q$.
The UV field theory displays $Z_N$ and $Z_{N-F}$ discrete subgroups of these  
symmetries. Viewing the curve asymptotically and as $\Lambda
\rightarrow 0$
\begin{eqnarray}
z \rightarrow \infty & w=0 & t = v^{N-F} m_Q^{F-N} \nonumber\\
z \rightarrow 0 & v = 0 & t = \left( {1 \over w}\right)^N 
\Lambda^{b_0} m_Q^{F-N}
\end{eqnarray}
The $U(1)_v$ and $U(1)_w$ symmetries (allowing $m_Q$ to transform
spuriously but not $\Lambda$) are indeed asymptotically broken to 
$Z_N$ and $Z_{N-F}$ discrete subgroups. Other combinations of
the two $U(1)_R$ symmetries may also be identified in the asymptotic
curve. For example $U(1)_A$ symmetry is given by the rotations
\beq
\begin{array}{c|cccccc}
& v & w & t & z & \Lambda^{b_0} & m_Q \\
\hline
U(1)_A & -2 & 2 & 0 & -2 & 2F & -2 \\
\end{array}
\eeq

The N=1 theory behaves like supersymmetric Yang Mills theory below the
matter field mass scale and has $N$ degenerate vacua associated with the
spontaneous breaking of the low energy $Z_{N}$ symmetry. In the curve this
corresponds to the $N$ curves in which $\xi_n = \xi_0 exp(2\pi
in/N)$ (equivalently $\Lambda^{b_0}_n = \Lambda^{b_0}_0 exp(2\pi
in)$). In the UV these curves can be made equivalent
by a $Z_{N}$ transformation.

\section{Non-Supersymmetric Solutions}

Let us now consider switching on susy breaking in the configurations
through $a\neq 0$ or equivalently switching on $\epsilon$.

\subsection{Softly Broken N=1 SQCD}

We begin by switching on $\epsilon$ in the N=1 configuration.
The curve is now 
\beq v = z +{\bar{\epsilon}\over \bar{z}} 
\hspace{1cm} w = {\Lambda^{b_0/N} m_Q^{F/N}\over z}, \hspace{1cm} t =
z^{N}m_Q^{F-N}/(z-m_Q)^{F}, \hspace{1cm} x^9 = 4 \epsilon^{1/2} Re \ln~ z
\eeq

The D4 branes lie in the $x^6$ and $x^9$ directions. This
distortion does not break the $U(1)_v$ or $U(1)_w$ symmetries and so
from the discussion above we deduce that in the $R\rightarrow 0$ limit
we have introduced the supersymmetry breaking terms
\beq   
D \left( |q|^2 - |\tilde{q}|^2 \right)
\eeq
The $w$ plane
has also been rotated from its supersymmetric position as can be seen
from the non-holomorphic nature of the first equation. The gaugino in
these configurations will also therefore be massive (and break both
$U(1)_R$ symmetries). We may identify the
parameter $\epsilon$ with field theory parameters from its symmetry
charges. Since it is chargeless under both $U(1)$s we may only identify it
as a function of
\beq
m_{\lambda}^N \Lambda^{b_0} m_Q^F
\eeq

To complete the identification we note that the field theory retains a
$Z_F$ subgroup of $U(1)_A$ symmetry even after the inclusion of the soft
breaking terms. Requiring this property of the curve asymptotically
forces
\beq
\epsilon =  \left( m_{\lambda}^N \Lambda^{b_0} m_Q^F \right) ^{1/N}
\eeq

Asymptotically the curve is then
\begin{eqnarray}
z \rightarrow \infty & w=0 & t = v^{N-F} m_Q^{F-N} \nonumber\\
z \rightarrow 0 & v = {\bar{m}_{\lambda}  \bar{w}}  & 
t = \left( {1 \over \bar{v}}\right)^N 
\bar{m}_\lambda^N\Lambda^{b_0} m_Q^{F-N}
\end{eqnarray}
which posseses a $Z_F$ subgroup of the U(1) 
\beq
\begin{array}{c|cccccccc}
& v & w & t & z & \Lambda^{b_0} & m_Q & m_\lambda & D\\
\hline
U(1)_A & -2 & 2 & 0 & -2 & 2F & -2 & 0 & 0\\
\end{array}
\eeq

The curves with different $\Lambda_{n}^{b_0}$ are no longer
equivalent on scales at which $\epsilon$ can not be neglected since the
$Z_N$ symmetry is broken. 
This is the behaviour of the field theory described in
\cite{15} where one of the $N$ vacua of the N=1 model becomes the true
vacuum of the N=0 model, the others become metastable vacua. 
The softly broken field theory solutions
exhibit first order phase transitions at $\theta_{phys} = ({\rm odd})
\pi$  where two of the $N_c$ vacua of the SQCD theory become degenerate
minima of the model. CP symmetry is spontaneously broken by the two
minima.
This behaviour has been identified in \cite{11}
for non-supersymmetric M-theory curves describing softly broken SQCD with
$F=0$. To do so the authors of \cite{11} 
allowed the parameter equivalent to $a$ to take complex 
values which allows $x^9$ to take multiple values. It is not clear that
the resulting brane configuration still corresponds to the field theory.
We note here that similar behaviour may be observed for these
configurations too. For simplicity
set the phases of $m_Q$ and $m_\lambda$ to zero (this may in general
be ensured by making $U(1)_R$ rotations; the theta angle is then the
physical theta angle $\theta_{phys} = \theta_0 + N \theta_{\lambda} +
arg(det m_Q)$). Shifting $\theta$ from 0 to $\pi$ 
has the effect of shifting $\Lambda_{0}^{b_0/N} 
\rightarrow \Lambda_{0}^{b_0/N}
exp(i \pi/N)$ and $\Lambda_{N-1}^{b_0/N} \rightarrow \Lambda_{0}^{b_0/N} 
 exp(-i \pi/N)$. These two
curves are then identical upto complex conjugation (CP symmetry in the
field theory). We deduce that as in the field theory the two vacua are
degenerate and break CP symmetry. 

From the softly broken field theory \cite{12} we expect that 
after supersymmetry
breaking a quark condensate will form. The quantity $\Sigma = m_\lambda^{N/F}
\Lambda^{b_0/F}$
has the correct $U(1)_R$ charges to play this role. It's dimension may
be corrected by a function of the symmetry invariants $|m|$, $|\Lambda|$
and $|m_{\lambda}|$.
It is $\Sigma \neq 0$ that breaks the
$Z_F$ symmetry of the curve in the IR. Note that a shift of the bare
$\theta_0$ angle shifts $\Sigma \rightarrow \Sigma exp(i \theta_0/F)$.

\subsection{Decoupling All Super-Partners}

If we let $\xi \rightarrow 0$ for fixed $\epsilon$ then we eliminate
all R-chargeful parameters except $m_Q$. The gaugino condensate has
therefore switched off and there is no parameter which may play the role
of the gaugino mass. We conclude that the gaugino has been decoupled
from the field theory. The curve is now

\beq \label{curve}
v= z + { (\bar{m}_Q \bar{\Sigma})^{F/N}
\over \bar{z}}, \hspace{1cm} 
t = z^{N}m_Q^{F-N}/(z-m_Q)^{F},
\hspace{1cm} x^{9} = 4 (m_Q \Sigma)^{F/2N} Re \ln z
\eeq

The configuration in the $R\rightarrow 0$ limit describes two NS5 planes
in the $v$ direction connected by $N$ D4 branes with $F$ semi-infinite
D4 branes in the $x^6,x^9$ directions. We conclude from the absence of
an adjoint matter field and gaugino that the configuration is in fact
one where one NS5 has been reversed by a rotation relative to the usual
supersymmetric set up and is an anti-NS5 brane.  

We expect in the field theory that all superpartners 
will have been decoupled from the
theory since without a massless gaugino there is nothing to stabilize
the scalar masses and they will radiatively acquire masses of order the
UV cut off. The curve provides support for this interpretation. In the
$m_Q \rightarrow 0$ limit $t \sim z^{N-F}$ and the gauge group has been
broken to an $SU(N-F)$ subgroup. In the field theory with non-zero $D$
but $m_Q=0$ there is a moduli space. The gauge group may be broken
leaving a mass gap below the UV scale of the field theory. All particles
decouple at the cut off except the $SU(N-F)$ subgroup and there is
therefore no renormalization of the scalar mass. For small non-zero
$m_Q^2 < D$, as discussed above, the classical theory has a vacuum that
higgs the gauge group at scales below D. The scalar fields exist in the
field theory below the UV cut off and therefore their masses will be
renormalized typically growing to of order the UV scale and hence
greater than $D$. The quantum
theory would therefore not be expected to have a higgsing vacuum for any
$m_Q$. This is the behaviour of the curve for non-zero $m_Q$, leading
some support to the hypothesis that scalar masses are being generated
radiatively.

With the decoupling of the gaugino the $U(1)_v$ symmetry becomes
$U(1)_A$ as may be seen in (\ref{v}).
Asymptotically the curve posseses a $Z_{F}$ subgroup of the
axial symmetry of the quarks although explicitly broken by the quark masses.
In the interior of the curve the symmetry is additionally 
broken by the quantity
$\Sigma$. We conclude that a quark condensate has
formed breaking the chiral symmetry.

There are $F$ distinct curves where $\Sigma_{n}
= \Sigma_{0} exp(i 2 \pi n/F)$ which are identical asymptotically
upto a spurious $Z_F$ transformation (since the mass explicitly breaks
the $Z_F$ symmetry the curves are inequivalent). 
We can also observe a non-trivial spectral flow with changing theta if
we again allow $a$ complex values. 
Setting the phase of $m_Q$ zero (this can be arranged by a $U(1)_A$
transformation leaving the physical theta angle $\theta_{phys} =
\theta_0 + arg(detm_Q)$) and
shifting $\theta \rightarrow \pi$ shifts $\Sigma_{0} \rightarrow 
\Sigma_{0} exp{(i \pi/F)}$ and $\Sigma_{(F-1)} \rightarrow 
\Sigma_{0} exp{(-i \pi/F)}$.  The two curves are again interchanged
by complex conjugation. We deduce that for these values of $\theta$
there are two degenerate vacua which spontaneously break CP. This is
precisely the behaviour observed in the QCD chiral lagrangian with
changing $\theta$ angle \cite{16}.

We conclude that the M-theory strong coupling extension of the type IIA
string configuration describing QCD in the IR behaves qualitatively as
we would expect QCD to behave. \vspace{4cm}

\noindent {\large \bf Acknowledgements}: I am very grateful to the
Department of Physics and Astronomy at the University of Kentucky for
hospitality and to Al Shapere and Clifford Johnson for detailed
discussions as this work was begun. I am also indepted to Mick Schwetz
and Martin Schmaltz for their advice. This work was in part supported by
the Department of Energy under contract $\#$DE-FG02-91ER40676.

\newpage


\end{document}